\documentclass{pramana}

\usepackage{graphicx,amsmath,bm}

\begin{document}

\title{Test of Isospin Conservation in Thermal Neutron-induced Fission of $^{245}$Cm}


\author{Swati Garg\textsuperscript{*} \and Ashok Kumar Jain}
\affilOne{Department of Physics, Indian Institute of Technology Roorkee, Roorkee-247667, India}


\twocolumn[{

\maketitle

\corres{swat90.garg@gmail.com}


\begin{abstract}
We have, recently, shown that the general trends of partitionwise fission fragment mass distribution in heavy ion (HI) induced compound nuclear (CN) fission of heavy nuclei can be reproduced reasonably well by using the concept of isospin conservation, hence providing a direct evidence of isospin conservation in neutron-rich systems~\cite{jain, swati, jain1, swati1}. In this paper, we test the concept of isospin conservation to reproduce the fission fragment mass distribution emerging from thermal neutron-induced CN fission reaction, $^{245}$Cm($n^{th}$, f). As earlier, we use Kelson's conjectures~\cite{kelson} to assign isospin to neutron-rich fragments emitted in fission, which suggest the formation of fission fragments in Isobaric Analog states (IAS). We calculate the relative yields of neutron-rich fragments using the concept of isospin conservation and basic isospin algebra. The calculated results reproduce quite well the experimentally known partition wise mass distributions. This highlights the usefulness of isospin as an approximately good quantum number in neutron-rich nuclei. This also allows us to predict the fragment distribution of the most symmetric Cd-Cd partition and the heavier mass fragment distributions, both not measured so far. 
\end{abstract}



}]



\section{Introduction}
Isobaric spin or, isospin depicts different electromagnetic states of a particle such as a nucleon, and is a fundamental tool for studying various nuclear processes~\cite{heisenberg}. In nuclear physics, one generally assigns isospin projection $T_3= +1/2$ for neutron and $T_3= −1/2$ for proton, which are two states of a nucleon having total isospin $T=1/2$. Isospin behaves in the same way as spin and follows the SU(2) algebra for nucleons. In particle physics, isospin can have values other than 1/2 also according to the various sets of particles involved; for example for pions, isospin $T_3= +1$ for $\pi^+$, 0 for $\pi^0$ and −1 for $\pi^-$, together forming an isospin triplet for $T=1$. However, isospin can assume very large values in nuclei, particularly in heavy nuclei and neutron-rich systems, where $N>Z$.

An early review by Robson~\cite{robson} presented the details of isospin algebra and also the selection rules involving isospin quantum number. Generally, isospin is considered to be very useful for light nuclei as it is a conserved quantity there. It is also relatively easy to assign isospin to these nuclei~\cite{warner}. In heavy mass nuclei, the Coulomb interaction becomes large and isospin mixing is thought to be significant, suggesting that isospin is not a good and useful quantum number.

A very lucid and succinct discussion of the various aspects of isospin impurity in heavy nuclei was presented by Soper as early as 1969~\cite{soper}, who underlined that until 1961, hardly any physicist would have taken isospin seriously as a quantum number beyond $A=60$. These assumptions were soon questioned by the discovery of isobaric analogue states in ($p$, $n$) reactions on nuclei near $A=90$~\cite{anderson}. 

In our recent works, we have been focusing on heavy mass nuclei which are natural $N>Z$ systems. Lane and Soper in 1962~\cite{lane} obtained a very interesting and useful result by using perturbation method which indicates that mixing of ground state isospin with the states having one unit higher isospin value decreases as the neutron excess increases. They considered the nucleus to be made up of a ($N=Z$) core  and ($N-Z$)  excess neutrons and calculated the impurity generated by the Coulomb potential of protons. It was shown that the impurity decreases with neutron enrichment, making isospin nearly a good quantum number in neutron-rich systems. Sliv and Kharitonov in 1965~\cite{sliv} also calculated the isospin mixing of $T=T_0+1$ into the ground state having $T=T_0$ using Coulomb potential as a perturbation and eigen-functions of harmonic oscillator. He estimated an isospin impurity of about 2$\%$ for $^{16}$O which rises upto 7$\%$ on reaching $^{40}$Ca. The impurity, however, starts decreasing as we move towards the heavier nuclei with $N>Z$ along the $\beta$-stability line, eventually reducing to 2$\%$ for $^{208}$Pb. Bohr and Mottelson have discussed the role of isospin in heavy nuclei~\cite{bohr} and also calculated the isospin mixing using the hydrodynamical model and concluded that isospin mixing is indeed very small for neutron-rich nuclei as compared to $N=Z$ nuclei. Auerbach~\cite{auerbach} in his review has compared the results of isospin mixing obtained from various approaches like the shell model, hydrodynamical model, RPA etc. and concluded that isospin impurity continues to decrease with the increasing neutron excess.

These findings along with our earlier results on HI induced fusion-fission reactions have encouraged us to test the validity of isospin as a good quantum number in the relative yields of fission fragments in thermal neutron induced CN fission of heavy nuclei. We use the same methodology which has been discussed earlier in~\cite{swati, jain1, swati1}. We note that the availability of precise data, where partitionwise fragment mass distributions are known to the precision of one mass unit, is still very scarce. This, however, is a must to test the idea of isospin conservation.  

We may emphasize that this paper does not calculate the fission fragment distributions from the first principles; our calculations rather take important inputs from the experimental data and show that the idea of conservation of isospin can reproduce the fission fragment distribution rather well, notwithstanding the crucial role that the shell effects play. In this paper, we analyze the fission fragment data from the reaction $^{245}$Cm($n^{th}$, f) as reported by Rochman $\it{et}$ $\it{al.}$~\cite{rochman}. We show that the experimental data of light mass fragments, as available from Rochman $\it{et}$ $\it{al.}$~\cite{rochman}, may be reproduced reasonably well, confirming an approximate validity of isospin in heavy nuclei. This also allows us to predict the mass distribution for heavy mass fragments and for the most symmetric partition, Cd-Cd.
  
\section{Formalism}
In thermal neutron induced fission, a neutron is incident on a target $X$ to form the compound nucleus (CN) which further fissions into two fragments $F_1$ and $F_2$ with the emission of $n$ number of neutrons. 
\begin{eqnarray*}
neutron(\dfrac{1}{2},\dfrac{1}{2})+X(T_X,T_{3_{X}}) \rightarrow CN(T_{CN},T_{3_{CN}})\rightarrow \\ F_1(T_{F1},T_{3_{F1}})+F_2(T_{F2},T_{3_{F2}})+n
\end{eqnarray*}

In order to test the validity of the isospin as a good quantum number in the fission process, we use the same formalism as reported earlier in our works~\cite{swati, jain1, swati1}. Our formalism is divided into two parts; the first part is to assign isospin to all the constituents present in the reaction and the second part is to calculate the relative yields of fission fragments emitted in fission based on the assigned isospin and related algebra.

\subsection{Assignment of isospin}
We start by assigning isospin to CN. The incident neutron has an isospin $T=T_3=1/2$. The target nucleus, assumed to be in its ground state, has minimum possible value of isospin $T=T_{3_{X}}$, where $T_{3_{X}}=(N-Z)/2$. Therefore, isospin of the CN, $T_{CN}$, should lie between $\mid T_{3_{X}}-1/2\mid$ and $(T_{3_{X}} +1/2)$. The third component of isospin of CN obviously has the value, $T_{3_{CN}}=(T_{3_{X}} +1/2)$. Since, the third component of isospin can have a value either less than or equal to total isospin, $T_3 \leqslant T$, it implies that the only possible value for $T_{CN}=T_{3_{CN}}=(T_{3_{X}} +1/2)$. For example, in the present case of thermal neutron induced fission $^{245}$Cm($n_{th}$, f), the isospin of target is $T$($^{245}$Cm)= 26.5. The isospin of the CN, $^{246}$Cm, can have two possible values, $T_{CN}= 26$ and 27 whereas $T_{3_{CN}}=27$. Therefore, the CN has a unique possible value of isospin, $T_{CN}=T_{3_{CN}}=27$.

We, now, proceed to assign the isospin values to various fission fragments. Before proceeding further, we introduce an auxiliary concept of residual compound nucleus (RCN) which is formed after the emission of $n$ number of neutrons from CN. Here, we have assumed that all the neutrons are emitted in one go, and no distinction is being made between pre-scission and post-scission neutrons, an approximation that seems to work well for our purpose. This simplifies the problem of many body system to a two body system. The third component of isospin of RCN is, therefore, given by
\begin{equation*}
T_{3_{RCN}}=T_{3_{CN}}-n/2=T_{3_{F1}}+T_{3_{F2}}
\end{equation*} 
 
The isospin value of RCN, therefore, may have a range of values given by,
\begin{equation}
\mid T_{CN}-n/2 \mid \leq T_{RCN} \leq (T_{CN}+n/2)
\end{equation}

Alternatively, it should also satisfy,
\begin{equation}
\mid T_{F1}-T_{F2} \mid \leq T_{RCN} \leq (T_{F1}+T_{F2})
\end{equation}

\begin{figure*}
\centering
\includegraphics[height=0.4\textheight]{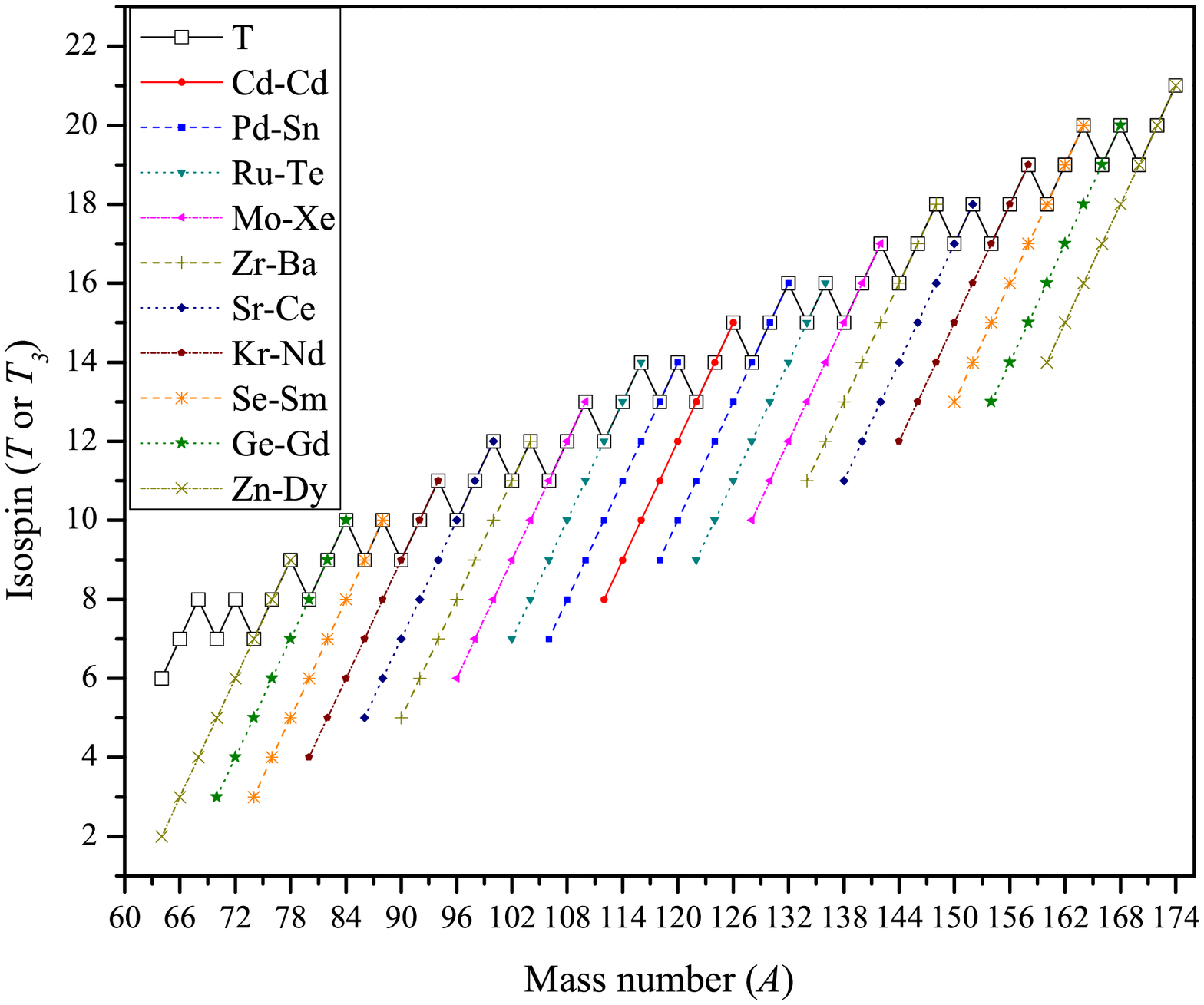}
\caption{\label{fig:tnt3}(Color online)  Assigned values of isospin $T$ or $T_3$ vs. mass number $A$ of the fission fragments emitted in the reaction $^{245}$Cm($n^{th}$, f). Open squares connected by the solid line show the isospin $T$ assigned to each mass number. Other symbols show the $T_3$ values for the fragments of different partitions. One particular type of symbol is used to connect $T_3$ values for the fragments of a distinct partition. Out of the two lines connecting the same type of symbols, the one on the right hand side is for the heavier and the one on the left hand side is for the lighter fragments.}
\end{figure*}

We fix the value of $T_{RCN}$ by using what we now term as Kelson's conjectures~\cite{kelson}. Kelson in 1969~\cite{kelson} considered the role of isospin and isobaric analogue states (IAS) in the fission phenomenon, and found it to be very useful and important in assigning the isospin values to the fission fragments. We have presented the detailed arguments in favor of these conjectures in our previous paper~\cite{swati1}. These two conjectures are: (i) As more and more neutrons are emitted in fission, the probability for the formation of highly excited states with $T>T_3$ increases. (ii) The fission fragments are preferably formed in IAS.

Using Kelson's first conjecture, we assign the isospin value of RCN as $T_{RCN}=T_{F1}+T_{F2}$ with the riding condition that it lies within the range given in Eq. (1). We then proceed to assign isospin values to the neutron-rich fission fragments for which we make use of Kelson's second conjecture. We choose three isobars corresponding to each mass number. These have same mass number but differ in $T_3$ values by two units, e.g. $T_3$, $T_3+2$ and $T_3+4$. As per Kelson's second conjecture, the fission fragments are preferably formed in IAS and, therefore, we consider the IAS of these three isobars for each mass number. We assign each mass number the isospin value $T$ which is maximum among the three $T_3$ values i.e. $T_3+4$, as this is the minimum value needed to generate all the members of any isobaric multiplet. For example, for mass number $A= 94$, we have $^{94}$Kr, $^{94}$Sr and $^{94}$Zr, which have $T_3$ values 11, 9 and 7 respectively. Therefore, we assign isospin $T= 11$ to $A= 94$ which is the maximum of the three $T_3$ values. We assign isospin value $T$ to each mass number in a similar fashion. The assignments so made are illustrated in Fig.~\ref{fig:tnt3}. 

We note that the experimental data are known only for the light mass fragments in a pair of fission fragments~\cite{rochman}. Therefore, to perform the complete calculations, we must also consider the corresponding heavy mass fragments. This gives us nine partitions namely Pd-Sn, Ru-Te, Mo-Xe, Zr-Ba, Sr-Ce, Kr-Nd, Se-Sm, Ge-Gd and Zn-Dy (from the symmetric combination to the most asymmetric combination). In addition, we also consider the most symmetric combination Cd-Cd to complete the three members for each mass number, although we do not have any experimental data on this partition. The assigned $T$ values for each mass number are shown by open squares in Fig.~\ref{fig:tnt3}. We can see from the figure that around the central partition Cd-Cd, isospin assignment is symmetric similar to what we obtained in our earlier work~\cite{swati1} for $^{208}$Pb($^{18}$O, f) and $^{238}$U($^{18}$O, f) reactions.

\subsection{Calculation of relative intensities of fission fragments in different partitions}
After the assignment of isospin values to all the fission fragments, we now proceed to calculate their relative yields. Cassen and Condon~\cite{cassen} introduced the isospin in wave functions so that the nuclear wave function should be anti-symmetric under space, spin and isospin coordinates. For our calculation, we consider only the isospin part of the total wave function involving isospin values of RCN and two fragments, $F_1$ and $F_2$. In a particular partition, for a given $n$-emission channel,
\begin{eqnarray}
\mid{T_{RCN},T_{3RCN}}\rangle_n = \langle{T_{F1}T_{F2}T_{3_{F1}}T_{3_{F2}} \mid T_{RCN}T_{3_{RCN}}}\rangle \nonumber \\ \mid{T_{F1},T_{3_{F1}}}\rangle \mid{T_{F2},T_{3_{F2}}}\rangle
\end{eqnarray}

where $n$ denotes a particular $n$-emission channel and the first part on the left hand side $\langle T_{F1}T_{F2}T_{3_{F1}}T_{3_{F2}} \mid T_{RCN}T_{3_{RCN}} \rangle$ represents the Clebsch-Gordon coefficient ($CGC$). The square of this $CGC$ is proportional to intensity of that particular pair of fragments. The yield of a particular fission fragment in a given $n$-emission channel for a particular partition may, therefore, be written as,
\begin{equation}
I_n = \langle{CGC}\rangle^2 = \langle{T_{F1}T_{F2}T_{3_{F1}}T_{3_{F2}} \mid T_{RCN}T_{3_{RCN}}}\rangle^2
\end{equation}

To calculate the total yield of a fragment, we take the sum of intensities from all the $n$-emission channels under consideration, $I = \sum_{n} I_n$. In Rochman $\it{et}$ $\it{al.}$~\cite{rochman}, there is no information about the weight factors of various $n$-emission channels. The average value of neutron multiplicity is 3.83 as reported by F. Gonnenwein~\cite{gonnenwein} in a talk. We perform two sets of calculations where we first consider 4$n$, 6$n$ and 8$n$ emission channels and then 2$n$, 4$n$ and 6$n$ emission channels.

\section{Results and Discussion}
\begin{figure*}
\begin{center} 
\includegraphics[width=14cm, height=15cm]{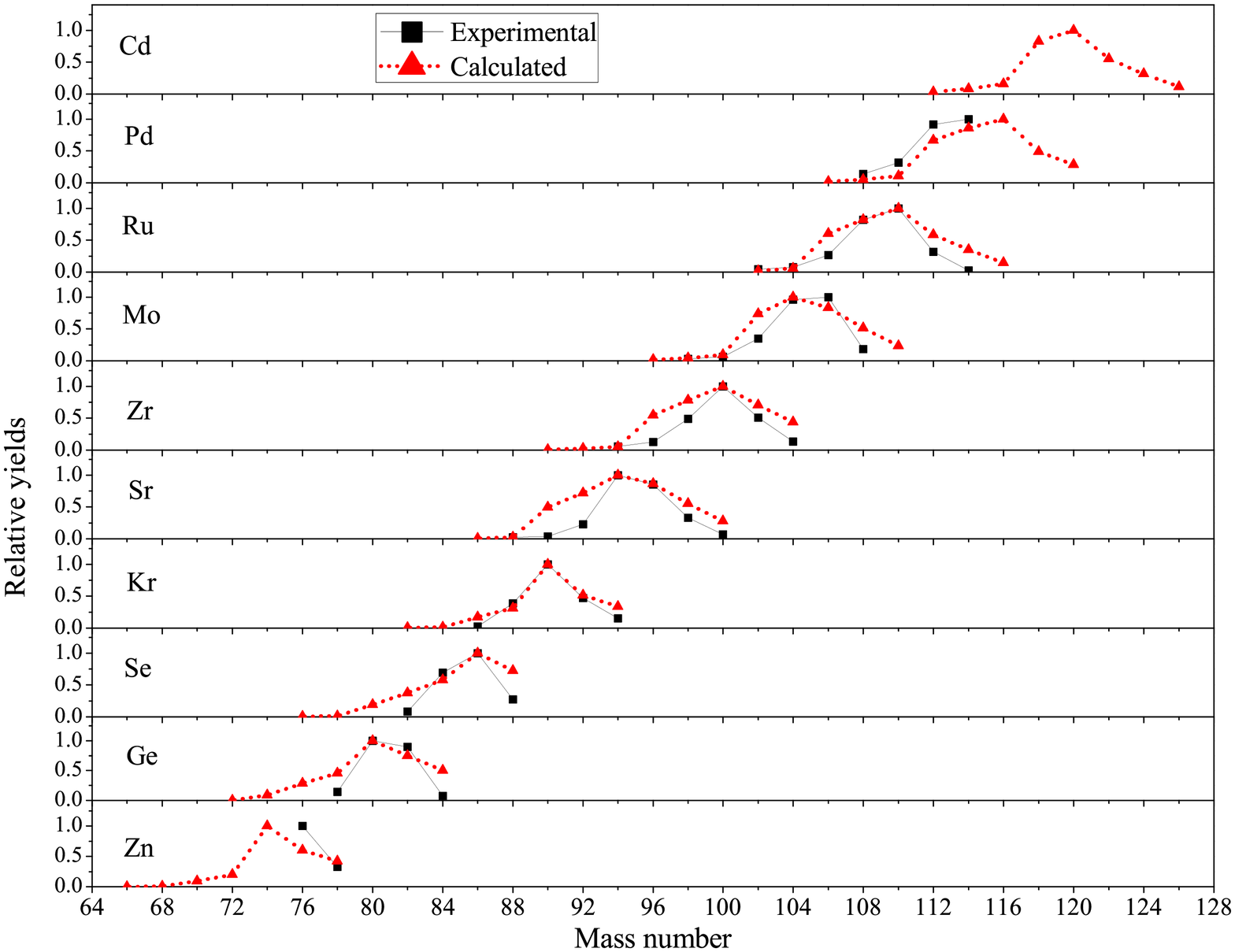}
\caption{\label{fig:rochman}(Color online) Comparison of the calculated and experimental relative yields of light mass fission fragments vs. mass number A for all the ten partitions formed in $^{245}$Cm($n^{th}$, f). Experimental data are taken from Rochman $\it{et}$ $\it{al.}$~\cite{rochman}. Note that there are no observed data for the Cd-Cd partition and also many additional fragments in all the partitions.}
\end{center}
\end{figure*}

We have performed the calculations for all the partitions using two combinations of $n$-emission channels, 4$n$, 6$n$, 8$n$ emission channels and 2$n$, 4$n$, 6$n$ emission channels. Since these calculations provide only relative yields, we must normalize the yields of all the fragments of a partition with respect to the maximum yield fragment for both the calculated yields and experimental data~\cite{rochman}. Comparison with the experimental data shows that for first six partitions, the 4$n$, 6$n$, 8$n$ emission channels combination works very well, and for the last four partitions, the 2$n$, 4$n$, 6$n$ emission channels combination gives quite good results. In Fig.~\ref{fig:rochman}, we plot our calculated relative yields with the experimental data for all the ten partitions, first six from 4$n$, 6$n$, 8$n$ emission channels and next four from 2$n$, 4$n$, 6$n$ emission channels.

We can see that our calculated results match with the experimental data fairly well in Fig.~\ref{fig:rochman}. There are some deviations which may be due to the shell effects, presence of isomers and side feeding of levels as discussed in Danu $\it{et}$ $\it{al.}$~\cite{danu, danu1}. The shell effects become prominent at closed shell configurations. The only closed shell configurations which may influence the data in the present case are $N=50$ and $N=82$. However, the total fission fragment distribution data available for light fragments only in Rochman $\it{et}$ $\it{al.}$~\cite{rochman}, do not display any significant dips due to shell closures of the same nature as seen by Danu $\it{et}$ $\it{al.}$~\cite{danu} and Bogachev $\it{et}$ $\it{al.}$~\cite{bogachev}. Also, there will be at least 5-10$\%$ error in the data. Even then, the overall agreement is quite good. These calculations are done without the inclusion of any weight factors as these are not known from the data.

\begin{figure*}
\begin{center} 
\includegraphics[width=14cm, height=15cm]{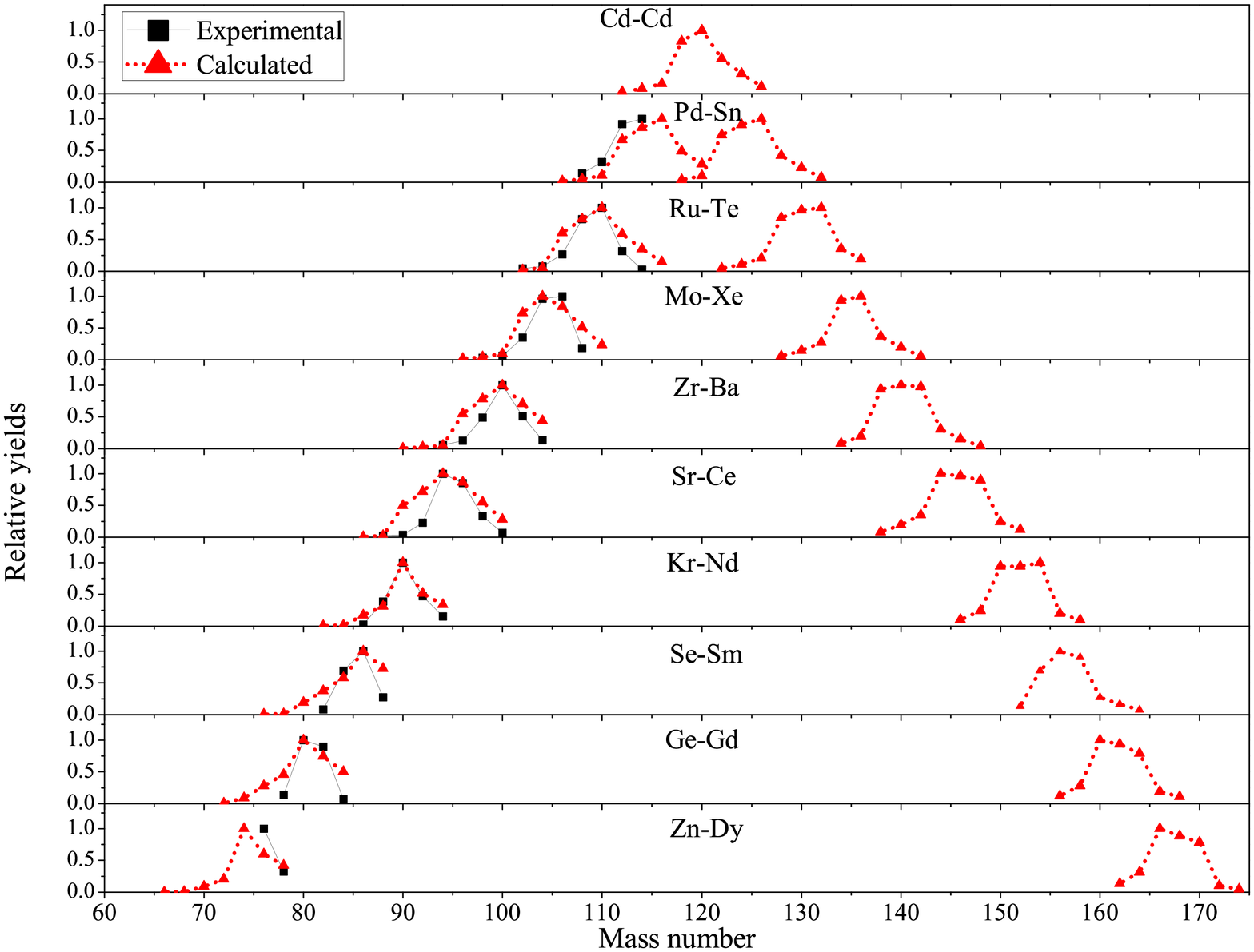}
\caption{\label{fig:rochmanlh}(Color online) Comparison of the calculated and experimental relative yields of both light and heavy mass fission fragments vs. mass number A for all the ten partitions formed in $^{245}$Cm ($n^{th}$, f). Experimental data are taken from Rochman $\it{et}$ $\it{al.}$~\cite{rochman}.}
\end{center}
\end{figure*}

In Fig.~\ref{fig:rochmanlh}, we plot the relative yields of both the light and the heavy mass fragments. This is an approximate prediction of the heavy mass fragments distribution. We have also predicted the possible distribution of the most symmetric partition i.e. Cd-Cd, which is plotted in Fig.~\ref{fig:rochman} and ~\ref{fig:rochmanlh}.

\section{Conclusion}
We calculate the partition-wise relative yields of fission fragments emitted in thermal neutron induced reaction $^{245}$Cm($n^{th}$, f) using the concept of conservation of isospin. For making the isospin assignments, we use Kelson's arguments who came up with the idea that the final fission fragments prefer to form in IAS with the emission of neutrons~\cite{kelson}. This idea helps us to assign isospin to all the fission fragments. The calculated results are in quite good agreement with the experimental data and also allow us to predict the mass distribution of heavy fragments not known so far. We also predict the fragment distribution of the Cd-Cd partition. We also note that there are many additional fragments in each partition for which no measurements are available. There are deviations at some points which may have many possible reasons like shell effects or presence of isomers.  Also, we expect that there will be at least 5-10$\%$ error in the experimental data. But, if we look at the complete description presented here, then we can say that isospin plays a very significant role in fission.  This also confirms Lane and Soper's idea of isospin purity in neutron-rich nuclei~\cite{lane}. The predictions made for the heavier fission fragments and the Cd-Cd partition stand as a challenge for the experimentalists. We believe that the modern fragment separators and/or, gamma ray tagging of fission fragments by using gamma ray arrays~\cite{danu, danu1, bogachev, banerjee} may be the right approach to identify them. These ideas may also help to predict more precisely the fission fragment mass distribution, if included in the theories of nuclear fission. 










\section*{Acknowledgement}
Support from Ministry of Human Resource Development (Government of India) to SG in the form of a fellowship is gratefully acknowledged.



\balance
\end{document}